\documentclass[showpacs,reprint,amsmath,amssymb,superscriptaddress,aps,prl]{revtex4-1}
\usepackage[percent]{overpic}
\usepackage{amsmath,amssymb}
\usepackage{graphicx}
\usepackage{float}
\usepackage{xcolor}
\begin{document}

\title{Extending time-domain ptychography to generalized phase-only transfer functions}

\author{Dirk-Mathys Spangenberg}\email{Corresponding author: dspan@sun.ac.za}
\affiliation{Laser Research Institute, Stellenbosch University, Private Bag X1, 7602 Matieland, South Africa}
\author{Erich Rohwer}
\affiliation{Laser Research Institute, Stellenbosch University, Private Bag X1, 7602 Matieland, South Africa}
\author{Michael Br\"ugmann}
\affiliation{Institute of Applied Physics, University of Bern, Sidlerstrasse 5, 3012 Bern, Switzerland}
\author{Thomas Feurer}
\affiliation{Institute of Applied Physics, University of Bern, Sidlerstrasse 5, 3012 Bern, Switzerland}

\date{\today}
\pacs{42.30.-d, 42.30.Rx, 42.65.Re}


\begin{abstract}
We extend the time-domain ptychographic iterative engine to generalized spectral phase-only transfer functions. The modified algorithm, i$^2$PIE, is described and its robustness is demonstrated by different numeric simulations. The concept is experimentally verified by reconstruction of a complex supercontinuum pulse from an all normal dispersion fiber.
\end{abstract}



\maketitle


    Recovery of coherent broadband signals, especially their phase, is a continuously developing field as detectors are often not fast enough to make direct measurements. Recently ptychography, a robust lens-less imaging technique developed to solve the phase problem in crystallography by Hoppe \cite{Hoppe1969}, has been migrated to the time domain \cite{Spangenberg2015a} by application of the ptychographic iterative engine (PIE) \cite{Faulkner2005} to time domain equivalent problems. Time-domain ptychography requires the measurement of a spectrum resulting from the product of two coherent signals, the unknown object and a time delayed probe signal. This is done for a number of time delays resulting in a sequence of spectra which is referred to as a spectrogram. In the most fundamental version the spectrogram is fed to the ptychographic iterative engine which reconstructs the unknown object signal given that the probe signal is known. Refined codes, e.g., the extended ptychographic iterative engine (ePIE), make use of redundancy in the spectrogram to also reconstruct the probe signal \cite{Lucchini2015}. More recently, a new modality, i.e., the implicit ptychographic iterative engine (iPIE), was introduced \cite{Spangenberg2015b,Spangenberg2016a}. In iPIE the spectrogram is generated from the product of an unknown object with a probe signal which is derived from the object by application of a linear spectral transfer function. Even though time-domain ptychography can be applied to all coherent broadband signals irrespective of carrier frequency, experiments published up to now focused on ultrafast broadband laser pulses. For example, PIE has been shown to reliably reconstruct unknown ultrafast pulses from corresponding spectrograms or cross-correlation frequency resolved optical gating (XFROG) traces \cite{Spangenberg2015a,Heidt2016}. 

In this work, we extend ptychography to reconstruct unknown object signals entirely without a probe signal, but by application of different families of spectral phase-only transfer functions. We call the scheme i$^2$PIE since we measure the \emph{square} of a signal which is the result of applying families of known transfer functions, i.e., \emph{intrinsic knowledge}, to the object signal. The new i$^2$PIE scheme has the potential to simplify ultrafast pulse reconstruction as no probe pulse is required. Instead, it analyzes spectra which come from collinear second harmonic generation of phase modulated object pulses. Thus, possible experimental arrangements are similar to those used in multiphoton intrapulse interference phase scan (MIIPS) \cite{Lozovoy2004a}, interferometric frequency resolved optical gating (iFROG) \cite{iFROG,Galler2008}, shaper assisted collinear spectral phase interferometry for direct electric field reconstruction (SPIDER) \cite{SPIDER} or the dispersion scan (D-Scan) method \cite{dscan}. Experimentally, we implement the method using a 4f-shaper with an SLM to apply selected spectral transfer functions and reconstruct a complex supercontinuum pulse from an all normal dispersion fiber.


The i$^2$PIE algorithm takes a spectrogram as input. The spectrogram $S(\Omega,n)$, consisting of $n$ measured spectra $S_n(\Omega)$, is recorded by applying each transfer function $H_n(\Omega)$ from a set of known spectral phase-only transfer functions $H(\Omega,n)$ sequentially to the unknown object signal and recording the resultant second harmonic spectrum. Here $\Omega = \omega-\omega_0$ is defined relative to the carrier frequency $\omega_0$. More formally, for each transfer function in a set, the product of the transfer function $H_n(\Omega)$ with the object pulse $E_\mathrm{in}(\Omega)$,

\begin{equation}
o_n(\Omega) = E_\mathrm{in}(\Omega) H_n(\Omega),
\end{equation}

is sent into a nonlinear mixer and the resultant spectrally resolved second harmonic intensity is recorded,

\begin{equation}
S_n(\Omega) = \left| \mathcal{F}\left\{ o_n^2(t) \right\} \right|^2,
\end{equation}

where $\mathcal{F}$ denotes the Fourier transformation. From such a spectrogram the unknown object signal $E_\textrm{in}$ can be reconstructed using the i$^2$PIE algorithm as follows. An initial guess is made for the object signal $E_\textrm{in}'(\Omega)$ which defines the modulated signal $o_n(\Omega)$ based on the corresponding transfer function,

\begin{equation}
o_n(\Omega) = E_\mathrm{in}'(\Omega) H_n(\Omega).
\label{eq_square}
\end{equation}

Assuming perfect phase matching over the entire spectral bandwidth, the second harmonic signal is

\begin{equation}
g_n(t) = o_n^2(t). 
\end{equation}

This second harmonic signal is used to calculate an updated field $g_n'$ by replacing the current estimated amplitude with the measured amplitude from the corresponding spectrum $S_n(\Omega)$, i.e.,

\begin{equation}
g_n'(\Omega) = \sqrt{S_n(\Omega)} \exp[ \textrm{i} \arg(g_n(\Omega)) ].
\end{equation}

Now the modulated signal is updated following standard ptychographic recipe

\begin{equation}
o_n'(t) = o_n(t) + \beta U_n(t) \left[g_n'(t) - g_n(t) \right]
\end{equation}

where

\begin{equation}
U(t) = \frac{|o_n(t)|}{\mathrm{max}\left(|o_n(t)|\right)} \; \frac{o_n^*(t)}{|o_n(t)^2|+\alpha}.
\label{eq_soft_div}
\end{equation}

We use a constant weight $\beta \in [0 \ldots 1]$ and $\alpha < 1$. The last step, unique to the i$^2$PIE algorithm, is to update the current estimate of the object signal $E_\mathrm{in}'$,

\begin{equation}
E_\textrm{in}'(\Omega) = o_n'(\Omega) H_n^*(\Omega),
\end{equation}

using the intrinsic knowledge of the transfer function used and the current updated second harmonic signal $o_n'$. The procedure is repeated for all recorded spectra multiple times until the object signal $E_\textrm{in}$ is sufficiently reconstructed.


As with other variants of PIE there is a redundancy of information requirement. This is achieved by having a sufficient number of transfer functions and sensible choices for the specific type of transfer functions. Here, we organize transfer functions into families with the same basis function and we show how one can calculate sensible boundary values for the free parameters of a transfer function family. More formally, to reconstruct the slowly varying envelope of an unknown object signal, using i$^2$PIE, 

\begin{equation}
E_\mathrm{in}(\Omega) = A(\Omega) \; \mathrm{e}^{\mathrm{i} \psi(\Omega)}
\end{equation}

a family of phase-only transfer functions $[\psi_n(\Omega)]$ with $n \in [1 \ldots N]$ is chosen. We restrict ourselves to families that can be characterized by only a few parameters and we will discuss two examples, i.e. families of polynomial and sinusoidal phases.  

The parameter boundaries of the chosen transfer function are either given by the experimental setup, or fundamentally, by discrete sampling theory. In the latter case they are just as easily obtainable for a measurement as they are for the simulated signals. Besides the transfer functions, we assume to know the spectral resolution $\Delta\Omega$ of the spectrometer and the spectrum of the unknown object signal $I(\Omega) = |A(\Omega)|^2$. From the spectrometer resolution we calculate the total time window, i.e. $T = 2\pi/\Delta\Omega$. Further, we assume that the maximum applied transfer function $\psi_N(\Omega)$ dominates the total phase, i.e. $\psi_\mathrm{tot}(\Omega) = \psi(\Omega) + \psi_N(\Omega) \approx \psi_N(\Omega)$. With this we can estimate the object signal duration after applying the maximum transfer function $\psi_N(\Omega)$ to

\begin{equation}
\label{eq_sigt}
\sigma_t^2 = \frac{1}{2\pi} \int\limits_{-\infty}^\infty \mathrm{d}\Omega \; \left\{ \left( \frac{\partial A(\Omega)}{\partial\Omega} \right)^2
+ \left[ \left( \frac{\partial \psi_N(\Omega)}{\partial\Omega} + \overline{t} \right) A(\Omega) \right]^2 \right\}
\end{equation}

with $\overline{t}$, the first moment of the temporal intensity. While the first term on the right hand side represents the bandwidth limited duration

\begin{equation}
\sigma_0^2 \doteq \frac{1}{2\pi} \int \mathrm{d}\Omega \; \left( \frac{\partial A(\Omega)}{\partial\Omega} \right)^2,
\end{equation}

the second term describes signal broadening due to the applied phase modulation. Hereafter, we assume $\overline{t}=0$ which is approximately true for most cases discussed here. We determine the parameter boundaries of $\psi_N(\Omega)$ by restricting the duration of the modulated object signal to a fraction $\gamma$ of the total time window, i.e. $\gamma T$. Therefore, broadening due to $\psi_N(\Omega)$ should at most be equal to

\begin{equation}
\sigma_\psi = \sqrt{\gamma^2 T^2 - \sigma_0^2}
\end{equation}

Consider the two families of transfer functions discussed hereafter. First, the family of polynomial phases

\begin{equation}
\psi(\Omega) = \pm q \, \Omega^k
\end{equation}

with parameter $q$ and constant order $k \geq 2$. With eq.~(\ref{eq_sigt}) we find for the maximum allowed $k$-th order phase

\begin{equation}
\label{eq_qmax}
q_\mathrm{max} = \pm \sqrt{\frac{\sigma_\psi^2}{\frac{k^2}{2\pi} \int \mathrm{d}\Omega \; \Omega^{2(k-1)} I(\Omega)}}
\end{equation}

which can be easily calculated knowing $\gamma T$ and the object spectrum $I(\Omega)$. Second, we consider the family of sinusoidal phase functions

\begin{equation}
\psi(\Omega) = a \cos(\Omega\tau+\phi)
\end{equation}

The members of this family are parameterized through amplitude $a$, frequency $\tau$ and phase $\phi$. With eq.~(\ref{eq_sigt}) we find

\begin{eqnarray}
\nonumber
a_\mathrm{max} \tau_\mathrm{max} & = & \sqrt{\frac{2 \sigma_\psi^2}{G(0) - \Re\{ G(2\tau_\mathrm{max}) \mathrm{e}^{2 \mathrm{i} \phi} \}}} \\
\label{eq_amax}
& \approx & \sqrt{\frac{2 \sigma_\psi^2}{G(0)}}
\end{eqnarray}

with the Fourier transform of the spectral intensity

\begin{equation}
G(t) \doteq \frac{1}{2\pi} \int \mathrm{d}\Omega \; I(\Omega) \; \mathrm{e}^{\mathrm{i} \Omega t}.
\end{equation}

Typically, the average spectral intensity $G(0)$ is larger than $\Re\{ G(2\tau_\mathrm{max}) \mathrm{e}^{2 \mathrm{i} \phi} \}$ and the approximate expression~(\ref{eq_amax}) can be readily used. Also note that the approximate expression is independent of the phase $\phi$. That is, we fix either $a_\mathrm{max}$ or $\tau_\mathrm{max}$ and use eq~(\ref{eq_amax}) to calculate the other.


We evaluate the i$^2$PIE algorithm by reconstruction of a set of random laser pulses. The pulses are used to numerically calculate input spectrograms based on a transfer function family, which are fed to i$^2$PIE. In each case we analyze how well the chosen transfer function family performs by testing its performance against the set of random object pulses. For each object pulse we calculate the root mean square (rms) error between input and reconstructed spectrogram. Reconstructions where $\log_{10}(\textrm{rms})<-3.5$ are considered successful and where $\log_{10}(\textrm{rms}) \geq -3.5$ are considered unsuccessful.

\begin{figure}[htbp]
\centering
\includegraphics[width=\columnwidth]{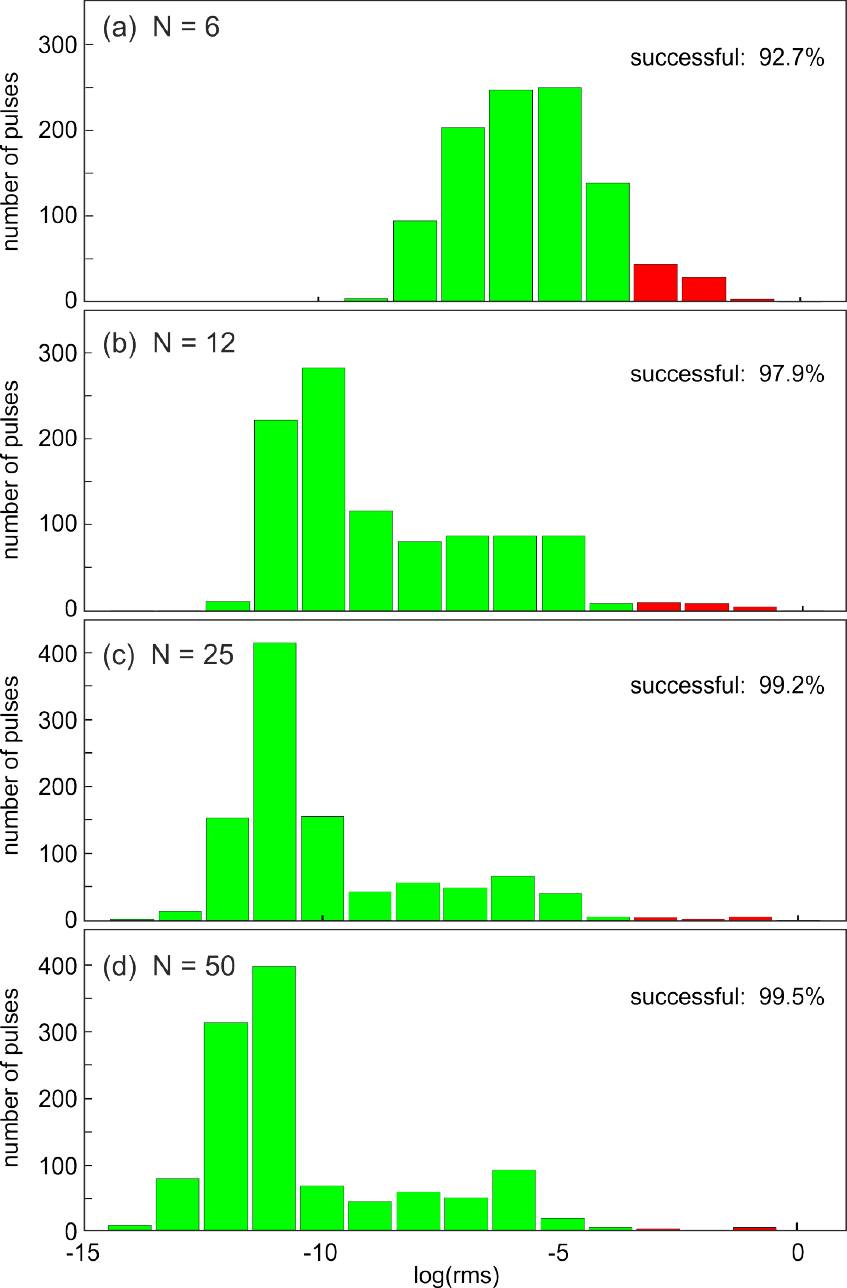}
\caption{Histogram of the logarithmic rms error of reconstructions of the random pulse set for a family of quadratic phase transfer functions with $N=6$ (a), $N=12$ (b), $N=25$ (c) and $N=50$ (d) members. Green bars indicate successful and red bars unsuccessful reconstructions.}
\label{figQ}
\end{figure}

The random object set consists of 1000 object pulses. Their randomly shaped spectrum is centered around 800~nm with a spectral bandwidth between 2~nm and 20~nm. The spectral phase is either, in the first case, polynomial up to fourth order with random coefficients and, in the second case, sinusoidal with random amplitude, frequency and phase. Each object pulse is generated for a temporal window of 8~ps, which corresponds to a spectral resolution of 0.27~nm at 800~nm, and on a grid of 1024 samples. In all reconstructions we used $\alpha=0.0001$ and $\beta=0.3$ and each reconstruction started with an initial guess of a 200~fs Fourier limited Gaussian pulse. We set $\gamma=1/8$ and used $N$ transfer functions in all families. For every object pulse we calculate the parameter boundaries of the respective transfer functions from $\gamma T$ and the fundamental spectrum according to eq.~(\ref{eq_qmax}) and (\ref{eq_amax}). We sequentially applied the i$^2$PIE update for the entire family of transfer functions in a set and repeat the process 500 times before the rms error was evaluated. 

First, we start with the family of quadratic phase transfer functions ($k=2$). The individual members are characterized by $q_n = (n-N/2-1) q_\mathrm{max}$. Shown in Fig.~\ref{figQ} are histograms of the logarithmic rms error of all reconstructions with $N=6, 12, 25, 50$. We find that with as little as six transfer functions the method successfully reconstructs 92.7\% of all objects. The success rate increases to close to 100\% for $N$ as large as 50 and the mean of the rms error decreases by several orders of magnitude. We find similar results for $k=3,4$.

\begin{figure}[htbp]
\centering
\includegraphics[width=\columnwidth]{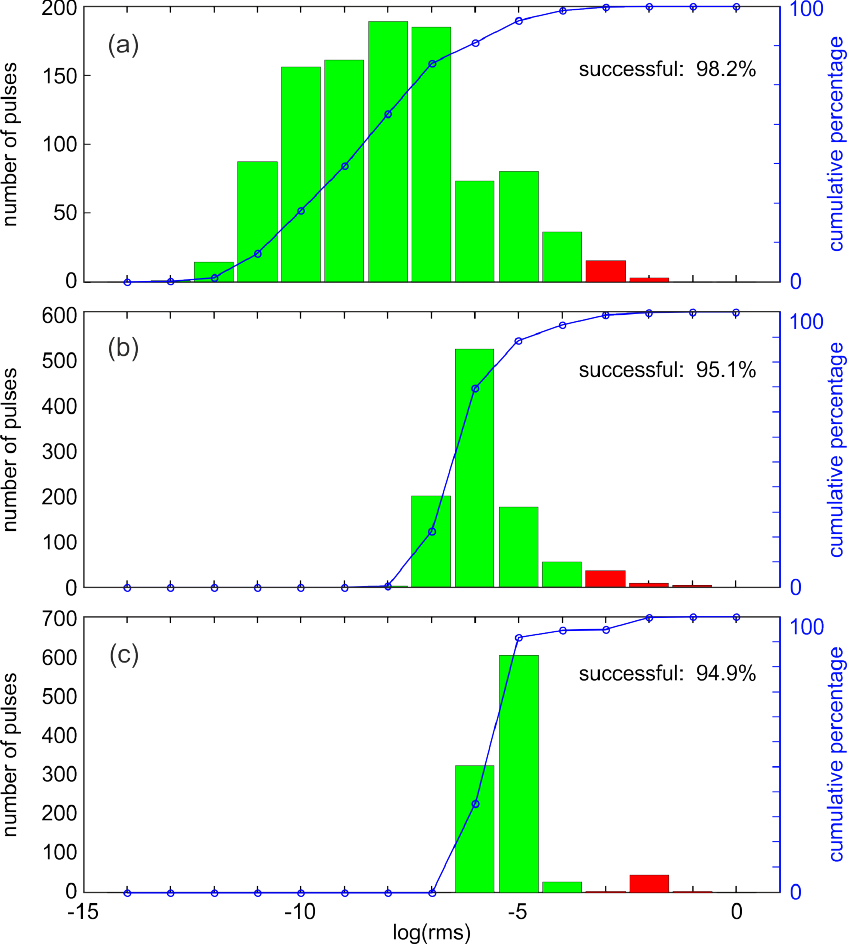}
\caption{Histogram of the logarithm of the rms error for three families of sinusoidal phase transfer functions. Green bars indicate successful and red bars unsuccessful reconstructions. The blue curve shows the cumulative percentage of successful reconstructions. (a) Varying $\phi$ from 0 to $2\pi$ for $\tau=300$~fs and $a$ from equation~(\ref{eq_amax}). (b) Fixing $\phi=0$, $\tau=300$~fs and varying $a$ within the limits calculated with equation~(\ref{eq_amax}). (c) Fixing $\phi=0$, $a=2.7$ and varying $\tau$ within the limits calculated with eq.~(\ref{eq_amax}).}
\label{fig1}
\end{figure}

Next we consider the families of sinusoidal phase transfer functions. Three families can be defined based on varying the parameters $a$, $\tau$ and $\phi$, respectively. First, we arbitrarily set $\tau=300$~fs and calculate the corresponding amplitude for every object pulse using equation~(\ref{eq_amax}). Then we vary $\phi_n$ between 0 and $\phi_N = 2\pi$ in $N$ equidistant steps. Second, we arbitrarily set $\phi=0$ and $\tau=300$~fs, and use equation~(\ref{eq_amax}) to calculate the maximum amplitude for every object pulse. The individual transfer functions then have amplitudes of $a_n = (n-N/2-1) a_\mathrm{max}$. Finally, we arbitrarily set $\phi=0$ and $a=2.7$, calculate $\tau_\mathrm{max}$ for every object pulse and vary the frequency according to $\tau_n = \tau_\mathrm{max} n/N$. Shown in Fig.~\ref{fig1} are histograms of the logarithmic rms errors for the three families of sinusoidal phase transfer function. The percentage of successful reconstructions is found to be 95\% and higher.


In our lab the broadband object pulse to be characterized is generated by sending a 800~nm seed pulse with 80~fs duration at 80~MHz repetition rate from a Ti:Sapphire oscillator into an all-normal dispersion (ANDi) photonic crystal fiber. The fiber output is then compressed by 48 bounces on a chirped mirror with 160~fs$^2$ compression per bounce. The resulting pulse serves as the object pulse $E_\mathrm{in}$. A 4f-shaper with a Jenoptic 640d spatial light modulator is used to sequentially apply all transfer functions $H_n(\Omega)$ from a specific family to the object pulse. The output from the 4f-shaper is focused onto a 20~$\mu$m thick BBO crystal by a 0.9~NA objective after which the frequency doubled light is collected and focused into an AvaSpec-3648 spectrometer with a resolution of 3.92~THz at 400~nm. The set of recorded second harmonic spectra is stored in a spectrogram $S_n(\Omega)$. 

\begin{figure}[ht]
\centering
\includegraphics[width=\columnwidth]{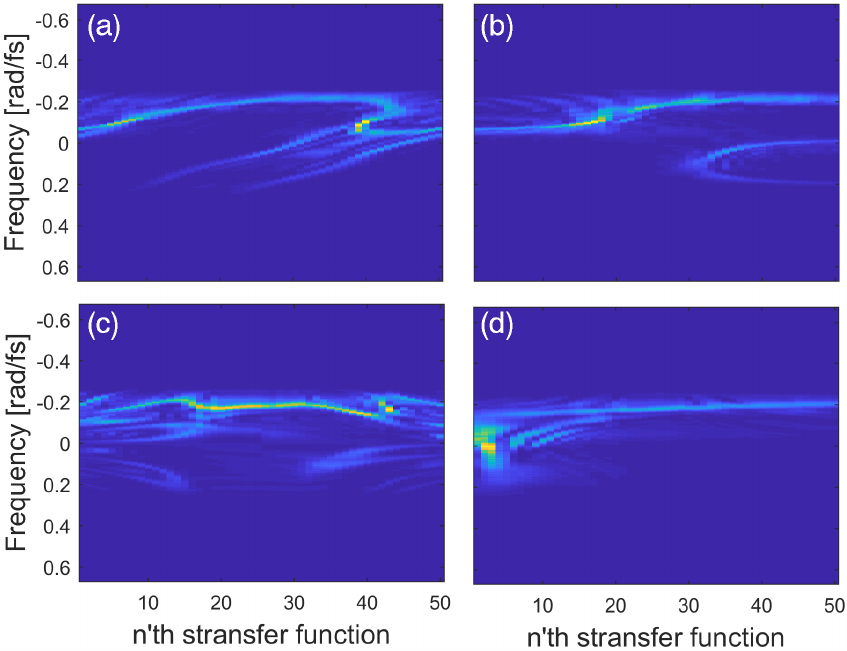}
\caption{Measured spectrogram for scanning (a) the quadratic phase between $-2250$~fs$^2$ and $2250$~fs$^2$, (b) the sinusoidal phase $\phi$ between $-\pi$ and $\pi$ with $a=15$ and $\tau=25$, (c) the sinusoidal amplitude $a$ between $-30$ and $30$ with $\phi=0$ and $\tau=25$~fs, and (d) the sinusoidal frequency $\tau$ between $-100$~fs and $100$~fs with $\phi=0$ and $a=\pi$.}
\label{fig_results}
\end{figure}

We took measurements based on families of quadratic and sinusoidal phase transfer functions where we varied the respective parameters as discussed in the simulation section. In Fig.~\ref{fig_results} the measured spectrograms are shown for the different cases when (a) the quadratic phase is varied between $-2250$~fs$^2$ and $2250$~fs$^2$, (b) the sinusoidal phase $\phi$ between $-\pi$ and $\pi$ with $a=15$ and $\tau=25$, (c) the sinusoidal amplitude $a$ between $-30$ and $30$ with $\phi=0$ and $\tau=25$~fs, and (d) the sinusoidal frequency $\tau$ between $-100$~fs and $100$~fs with $\phi=0$ and $a=\pi$. The spectrograms are further used to retrieve amplitude and phase of the object pulse.

\begin{figure}[ht]
\centering
\includegraphics[width=\columnwidth]{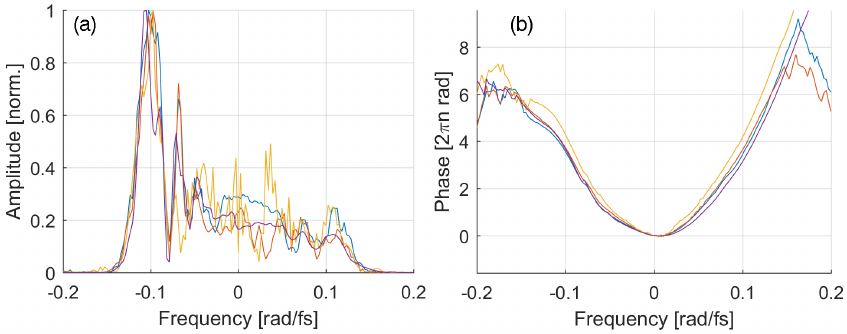}
\caption{The reconstructed spectral intensity is shown in (a) and the phase in (b). Purple: quadratic phase scan; blue: $\phi$ scan; red: amplitude $a$ scan; yellow: frequency $\tau$ scan.}
\label{fig_amp_phase}
\end{figure}

In Fig.~\ref{fig_amp_phase}(a) we plot the reconstructed spectral intensities for all four families of spectral transfer functions on top of each other and in Fig.~\ref{fig_amp_phase}(b) the respective reconstructed spectral phases. We find reasonable agreement in the reconstructed spectral amplitudes and excellent agreement in the retrieved phases in regions of nonzero amplitude.


In summary, we have demonstrated that the i$^2$PIE algorithm can reconstruct amplitude and phase of an unknown object signal from a measured second harmonic spectrogram recorded by applying different families of phase-only spectral transfer functions with excellent results. In principle, the choice of family is arbitrary and we derive a formalism that allows to calculate the scan limits from only the spectral resolution of the spectrometer and the spectral intensity of the object pulse.

\section*{Acknowledgments}

This research was funded in part through the Swiss National Science Foundation (Grant Number: 200020-178812/1) as well as the National Research Foundation of South Africa (Grant Number: 47793).

\end{document}